\documentclass[twocolumn,amsmath,amssymb]{revtex4}
\usepackage{graphicx}
\usepackage{booktabs}

\begin{document}

\title{Details of soft particle clogging in two-dimensional hoppers}

\author{Ran Tao}
\author{Madelyn Wilson}
\author{Eric R. Weeks}
\email{erweeks@emory.edu}
\affiliation{Department of Physics, Emory University,
Atlanta, GA 30322, USA}

\date{\today}

\begin{abstract}
We study the outflow of soft particles through quasi-two-dimensional hoppers with both experiments and simulations. The experiments utilize spheres made with hydrogel, silicone rubber, and glass.  The hopper chamber has an adjustable exit width and tilt angle (the latter to control the magnitude of gravitational forcing).  Our simulation mimics the experiments using purely two-dimensional soft particles with viscous interactions but no friction.  Results from both simulations and experiments demonstrate that clogging is easier for reduced gravitational force or stiffer particles.  For particles with low or no friction, the average number of particles in a clogging arch depends only on the ratio between hopper exit width and the mean particle diameter. In contrast, for the silicone rubber particles with larger frictional interactions, arches have more particles than the low friction cases.  Additionally, an analysis of the number of particles left in the hopper when clogging occurs provides evidence for a hydrostatic pressure effect that is relevant for the clogging of soft particles, but less so for the harder (glass) or frictional (silicone rubber) particles.
\end{abstract}

\maketitle

\section{Introduction}
\label{intro}

The hopper discharge of granular materials has been intensively studied due to its practical importance to industries, such as agriculture, architecture, and mining \cite{deming29,franklin55,brown58,fowler59,beverloo61,jenike67}.  Hoppers are containers with funnel-shaped bottoms where particles can flow out.  By adjusting the width of the hopper opening, the flow rate can be controlled \cite{beverloo61}.  However, at small opening widths, clogging can occur, in particular if the particles can form an arch that spans the width of the opening \cite{to01}.  This typically happens at a critical opening width of about 3 to 6 particle diameters \cite{deming29,brown58,beverloo61,nedderman82,sheldon10,aguirre10,wilson14}.  Even when the opening width is greater than the critical size, the flow rate fluctuates due to transient clogging events \cite{zuriguel14,harth20}, affecting the ability to smoothly dispense granular materials out of a hopper.  In everyday life, the flow of granular materials has been applied to the study of the movement of people during emergency evacuations \cite{helbing00,garcimartin17,hidalgo17}. There are already many prior studies on the flow and clogging of hard particles; however, the outflow of soft particles still lacks a comprehensive physical description.

Prior work showed the importance of softness to the clogging process.  Experiments showed that due to soft particles’ ability to deform, clogging only occurred for much smaller opening widths compared with results from the studies of hard particles \cite{bertho06,lumay15,hong17}, which significantly changed the flow rate \cite{ashour17}.  Slightly above the critical opening size, there could be long-lived transient clogs that eventually unclogged \cite{harth20}.  For even larger opening sizes, it was found that the flow rate and internal velocity fields differ for soft particles compared to hard particles \cite{stannarius19,harth20,pongo21}.  Simulations of soft frictionless granular materials demonstrate that clogging is easier for stiffer particles or with weaker gravitational forces \cite{hong17}.  Prior experimental studies of soft particles mostly focused on hydrogel particles \cite{hong17,ashour17,stannarius19,harth20,pongo21} although one study also included oil droplets \cite{hong17} which, due to their easy ability to deform, were even harder to clog than hydrogel particles.

In this paper, we study clogging in the outflow of a hopper using a quasi-two-dimensional experiment with granular materials of varying softness, and simulations mimicking frictionless soft particles.  Our experiment uses glass particles, silicone rubber particles, and hydrogel particles, as shown in Fig.~\ref{ranpics}.  The hopper can be tilted relative to gravity, allowing us to adjust the driving force.  The choice of particles and tilt angles allow us to vary the particle effective stiffness by a factor of $10^4$.  For the harder particles (glass, silicone rubber; lower gravity) clogging is easier and occurs with larger opening widths; for the softer hydrogel particles the opposite is true.  The simulation results agree well with the experimental results.  Both the experiment and simulations show that the number of particles forming the arch is determined by the ratio between the opening width and the particle diameter regardless of the particle softness.  The sole exception is for the silicone rubber particles, which have a markedly higher friction coefficient leading to larger arches.  Finally, an examination of the number of particles in the hopper when a clog occurs reveals that the hydrostatic pressure of the soft particles causes clogging to be exponentially less likely when the hopper is full.  The exceptions are for the glass particles and silicone rubber particles, suggesting hardness and friction change the physics of soft particle clogging.

\section{Experimental Methods}

\begin{figure}
\includegraphics[width=8.0cm]{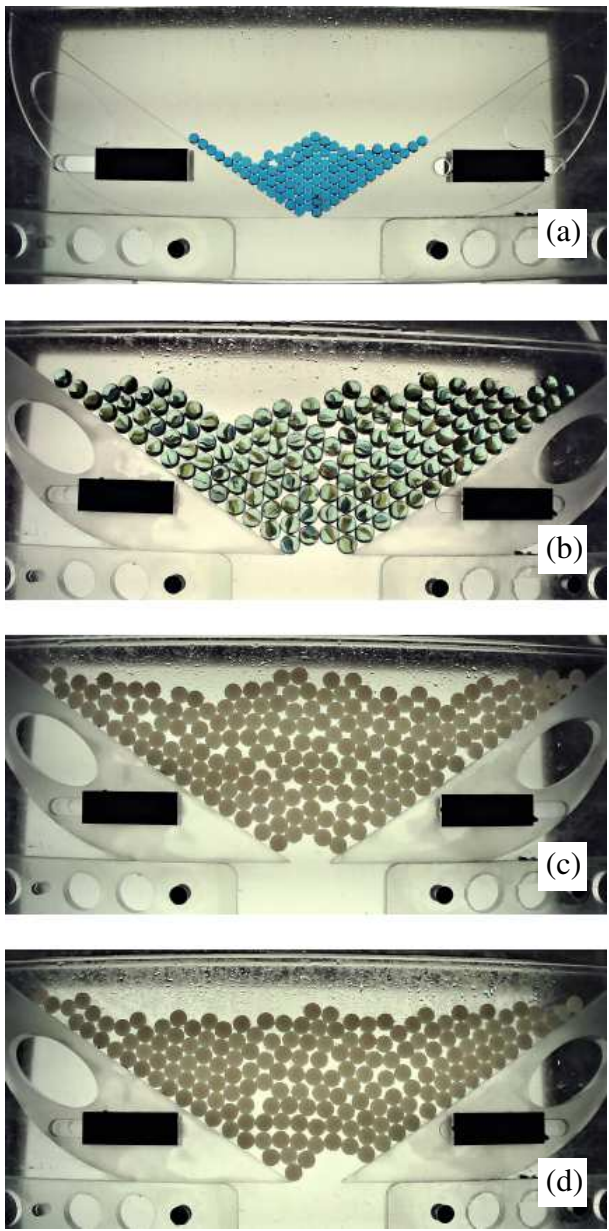}
\caption{\label{ranpics} Photograph of different types of particles in a clogged state, with the sample chamber fully vertical (maximum force of gravity). (a) Clogging of small hydrogel particles, showing a 3 particle arch.  The opening width is $w = 14.3$~mm $= 1.81d$ in terms of the mean particle diameter $d = 7.9$~mm. (b) Clogging of glass spheres, showing a 4 particles arch with $w = 51.5$~mm $= 3.32d$, $d = 15.5$~mm. (c) Clogging of silicone rubber particles, showing a 5 particle arch with $w = 30.6$~mm $= 2.19d$, $d = 14.0$~mm.  (d) Clogging of silicone rubber particles, showing a 6 particle arch with $w = 39.9$~mm $= 2.85d$.}
\end{figure}

The apparatus used in our experiment is the same as the one described in our group's prior work \cite{hong17}; we reprise the key details here.  The hopper has two movable sidewall blocks at 34$^\circ$ angles measured from the horizontal, pictured in Fig.~\ref{ranpics}.  Above and below the main hopper chamber there are two identical storage chambers.  To initialize the experiment, we place 200 particles in the upper storage chamber. A bottom metal plate inserted between the upper storage chamber and the hopper holds these particles. To begin the experiment, we rapidly remove the metal plate by hand allowing the particles to fall. Particles that fall through the hopper are collected by the storage chamber below the hopper. The two storage chambers are then swapped, moving the particles back to the top position, readying for the next trial.  If a clog occurs, the sidewall blocks are moved to let those particles drain out.  To ensure a reproducible opening width between trials, the hopper blocks are pushed against an inserted plastic spacer with the desired opening width and then locked into place.  The entire apparatus is mounted on a horizontal axis, so that we can vary the component of the gravitational force in the plane of the hopper by setting the tilt angle $\theta$ relative to the horizontal.  In practice, we vary the component of gravity in the hopper by a factor of 6.  At the smallest tilt angles, occasionally particles get stuck between the bottom of the upper storage container and the top of the hopper, on the seam between the two parts.  Accordingly, we avoid tilt angles where this problem occurs, which is the limiting factor on our ability to adjust the gravitational force.

We use three types of particles with varying softness in the clogging experiments:  hydrogel particles, silicone rubber particles, and glass particles. The physical properties of the particles are given in Table I.  The hydrogel particles are a polyacrylamide gel (blue water beads, purchased from AINOLWAY, amazon.com). When these hydrogel particles are dry, they are spheres with diameters around 3 mm and moderate polydispersity.  We use two sieves to constrain the dry particle diameter to be between 2.80 mm and 3.15 mm.  We then swell the hydrogel particles in salt water; by changing the concentration of salt, we can control the final diameter of the hydrogel particles.  Salt water with concentration 0.01 mole/L swells the hydrogel particles to a mean diameter of 13.8 mm, and a concentration of 0.5 mole/L results in a mean diameter of 7.9 mm.  Additionally, we use silicone rubber spheres with diameters 14.0 mm (purchased from Hebei Baorui Rubber Products Company, alibaba.com), and glass spheres with diameter 15.5 mm (marbles purchased from amazon.com).  The silicone rubber, glass particles, and large hydrogel particles are used in sample chambers with thickness 17.0 mm.  For the small hydrogel particles we adjust the thickness to 9.0 mm.  Table I lists these diameters along with their standard deviations.

To measure physical properties of our hydrogel and silicone rubber particles, we use a TA Instruments AR2000 rheometer with a parallel-plate geometry.  To measure the Young’s modulus we compress individual spheres and measure the normal force.  The resulting relation between the compression force and the displacement is well fit by the Hertzian force law, and provides us with the modified Young's modulus $E^* = E/(1-\nu^2)$ in terms of the Poisson ratio $\nu$ and Young's modulus $E$.  Later we will need $E^*$, so the data are listed in Table I.  The small and large hydrogel particles come from the same dry particles, so accordingly the larger elastic modulus of the smaller particles (swelled in high concentration salt water) is due to the higher polymer concentration of the smaller hydrogel particle.  The glass particles are too stiff to be measured in the rheometer, so the quoted modulus is an estimate (from https://www.engineeringtoolbox.com/young-modulus-d\_417.html); the main point is that the elastic modulus of glass is several orders of magnitude larger than the other particle types.

\begin{table}[th]
\begin{center}
\begin{tabular}{cccc}
\toprule
Particle & $d$ (mm) & $E^*$ (kPa) & $\mu$ \\
\hline
\midrule
large hydrogel    & $13.8\pm 0.6$ & $54.0\pm 6.4$ & $0.004\pm 0.002$ \\
small hydrogel    & $7.9\pm 0.2$  & $60.4\pm 4.9$ & $0.004\pm 0.002$ \\
silicone rubber    & $14.0\pm 0.1$ & $5100\pm 100$ & $0.4\pm 0.2$ \\
glass             & $15.5\pm 0.1$ & $(8\pm 2)\times 10^7$ & $0.009\pm 0.002$\\
\bottomrule
\hline
\end{tabular}
\end{center}
\caption{
The diameter $d$, Young's modulus $E$, and coefficient of sliding friction $\mu$ for each particle type.  The large hydrogel particles are made by swelling the dry particles in 0.01~M NaCl, while the small hydrogel particles use 0.5~M NaCl.  The diameters are listed as mean $\pm$ standard deviation.  The uncertainties of $E$ and $\mu$ are based on the reproducibility of our measurements.  The exception is the uncertainty of $E$ for the glass particles, which is based on differences in literature values.
}
\label{pj}
\end{table}

To measure the surface friction of the particles, we use the technique described in our group's prior work \cite{hong17}.  We place a pair of particles symmetrically a distance $R=1$~cm from the rheometer axis.  The particles are trapped in small wells made from glue and paper to prevent the particles from rolling. The parallel plate rheometer tool compresses the particles slightly with normal force $N$, and we then measure the torque $\tau$ required to rotate the rheometer tool.  The friction coefficient is then calculated from $\mu = \tau / 2 N R$.  The results depend somewhat on the rotation speed \cite{cuccia20}, so we have uncertainties of 50\% listed in Table I for all but glass (where the results vary less).  The main point made in Table I is that the silicone rubber particles have a friction coefficient about $50-100$ times larger than the other particle types.

To look for clogging, we load the hopper with 200 particles and allow them to flow through the hopper with a fixed opening width. We record whether the experiment clogs.  If a clog occurs, we wait at least a minute to confirm the particles are stationary; in practice any transient clogs last only a few seconds, in agreement with our group's prior experiments \cite{hong17} and observations by Harth {\it et al.}~\cite{harth20}.  Clogging probabilities are measured by repeating each condition at least 20 times.

\section{Simulation Methods}

In addition to the experiments, we also do simulations using the two-dimensional Durian Bubble Model \cite{durian95,tewari99} as modified in our group's prior work \cite{hong17}.  This model assumes strong viscous forces such that the velocity-dependent viscous forces are balanced by all other forces, and thus at each time step a differential equation is solved for the velocity rather than the acceleration.  This differential equation for each particle $i$ is:
\begin{equation}
\label{bubblemodel}
\sum_{j}[ \vec{F}^{\rm contact}_{ij} +
\vec{F}^{\rm visc}_{ij}(\vec{v_i},\vec{v_j})] +
\vec{F}^{\rm wall}_i +
\vec{F}^{\rm grav}_i +
\vec{F}^{\rm drag}_i(\vec{v_i}) = 0.
\end{equation}
Each soft particle has a radius $R_i$ and the contact force is zero if two particles do not overlap.  For overlapping particles $i$ and $j$, the repulsive contact force is given by
\begin{equation}
\label{repulsive}
\vec{F}_{ij}^{\rm contact} =
F_0 \Big[\frac{1}{|\vec{r}_i - \vec{r}_j|} -
\frac{1}{|R_i + R_j|} \Big] \vec{r}_{ij},
\end{equation}
based on their positions $\vec{r}$, defining their separation as $\vec{r}_{ij} = \vec{r}_j - \vec{r}_i$, and requiring $|\vec{r}_{ij}| < (R_i + R_j)$ for overlaps.  The viscous force is experienced by overlapping particles moving at different velocities and is given by $\vec{F}^{\rm visc}_{ij} = b(\vec{v}_j-\vec{v}_i)$.  The wall force is a contact force experienced by droplets which overlap the wall, treating the wall as a particle with $R_j=0$ in Eqn.~\ref{repulsive}.  The gravitational force is $F^{\rm grav}_i = - \rho g R_i^2 \hat{y}$, proportional to the particle area.  This model was inspired by experiments with oil droplets compressed between two parallel plates \cite{hong17}, so the final force is a drag force coming from these plates, $F^{\rm drag}_i = -c R_i^2 \vec{v}_i$.  We set $F_0=b=c=\rho=1$ and vary $g$ to influence the importance of particle softness.  The mean droplet radius $\langle R \rangle$ is set to be 1.  The unit of time is $b \langle R \rangle/F_0$, which is the time scale for two droplets to push apart, limited by inter-droplet viscous interactions.  With these parameter choices, the free-fall velocity of an isolated particle is $g$.  See Ref.~\cite{hong17} for further discussion about these parameter choices.

The specific geometry is matched to the experiment, with a hopper wedge angle of $34^\circ$.  We simulate 800 particles with a polydispersity of 0.1 (the polydispersity is the standard deviation of $R_i$ divided by $\langle R_i \rangle=1$).  The particles are initialized in random positions above the hopper exit, and then allowed to fall toward the exit.  Equation \ref{bubblemodel} is computed using the 4th order Runge-Kutta algorithm with a time step of 0.1, except for the simulations with $g=10^{-4}$ where a time step of 1.0 is used.  In some cases, the simulation ends with all of the particles falling out of the hopper, defining a situation without a clog.  In other cases, the simulation is ended when the maximum speed of all particles in the hopper falls below $10^{-10}$, defining a clog.  In practice, once the maximum velocity of the particles in the hopper is below $10^{-6}$, their velocities decay exponentially toward zero; the particles do not unclog \cite{hong17}.  Stated another way, we do occasionally observe long transient clogs where the particles have slight motions and eventually unclog, and these transients always have a maximum velocity of at least one particle above $10^{-6}$, allowing for the rearrangements necessary to unclog and returning to a flowing state.

\section{Results}

\subsection{Clogging Probability}

\begin{figure}[ht!]
\includegraphics[width=8cm]{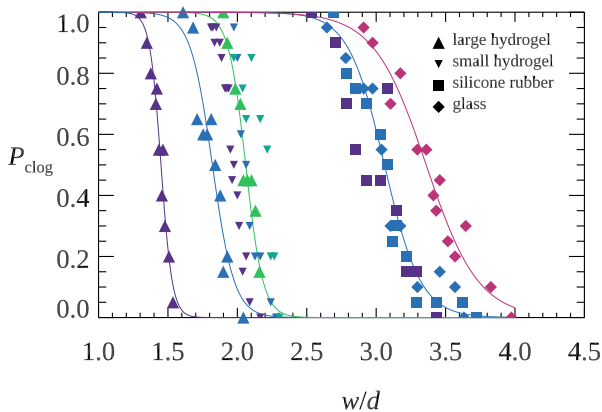}
\centering
\caption{\label{clogprob}
Experimental probability of clogging as a function of $w/d$, the ratio of the hopper exit width $w$ to the droplet diameter $d$. Data for different types of particles with the influence of gravity varied by setting different tilt angles. Different symbols represent different types of particles as indicated in the legend.  The different colors stand for different tilt angles $\theta$: $90^\circ$ (dark purple, largest influence of gravity), $50^\circ$ (dark blue), $35^\circ$ (light blue), $20^\circ$ (green), and $10^\circ$ (red, smallest influence of gravity).  The lines are sigmoidal fits to the large hydrogel data and the glass data:  $P_{\rm clog}(w/d) = \{1 + \exp[(w/d - a)/s]\}^{-1}$.
}
\end{figure}

Our first goal is to investigate the clogging probability. The experimental data are shown in Fig.~\ref{clogprob}.  For the experiment, we repeat each experimental condition 20 times and compute the clogging probability $P_{\rm clog}$ from the fraction of times that we observe clogging. We go through the same process for different types of particles, different values of opening width $w$, and different hopper tilt angles $\theta$.  Each set of symbols illustrates that $P_{\rm clog}$ decreases as we enlarge the hopper opening width for a fixed gravitational force.  Overall, these results are consistent with our group's prior experimental work with slightly different hydrogel particles \cite{hong17}.

% this is kingkong: /data/eric/hopper-to-dr/sigmoid_fitting/mkclog2

\par

In Fig.~\ref{clogprob} each family of symbol type indicates one type of particles. The influence of gravity is apparent: clogging is easier for reduced gravity, as signified by the curves shifting to the right in Fig.~\ref{clogprob} as the colors vary from dark purple (maximal gravity) to blue to red (minimal gravity). For large hydrogel particles, as gravity decreases by a factor of 3, the location where $P_{\rm clog}=1/2$ shifts from $w/d \approx 1.4$ to 2.1. Different types of particles behave differently in the clogging experiment: harder particles are more likely to clog, also signified by the curves shifting to the right. With a tilt angle $\theta=50^\circ$, the large hydrogel particles have $P_{\rm clog} \approx 1/2$ at $w/d \approx 1.8$, the small hydrogel particles have $P_{\rm clog} \approx 1/2$ at $w/d \approx 2.1$, and both the silicone rubber particles and glass particles have $P_{\rm clog} \approx 1/2$ at $w/d \approx 3.1$.

\begin{figure}
    \centering
    \includegraphics[width=8cm]{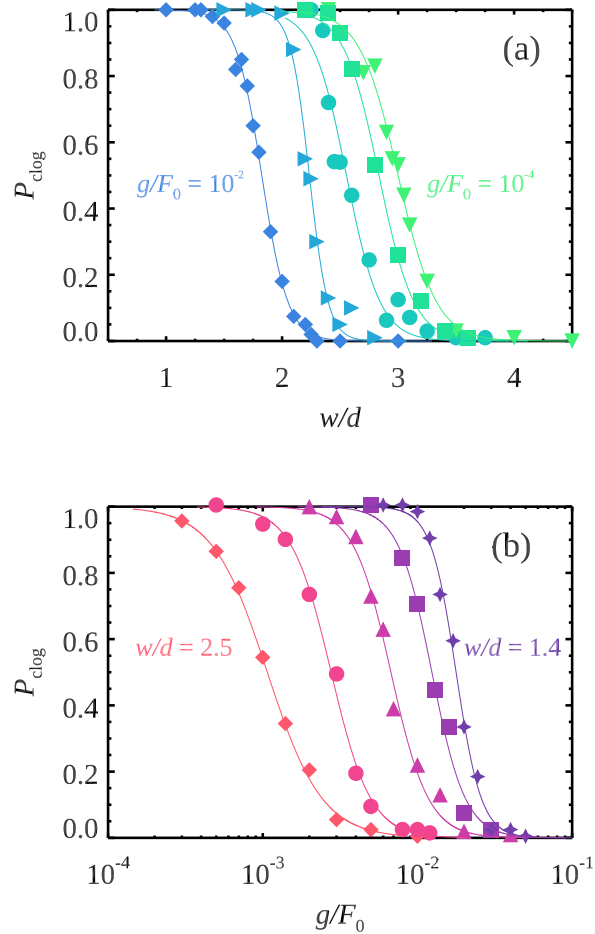}
    \caption{Clogging probability from the simulation.  (a) Clogging probability as a function of $w/d$ for fixed values of the driving force $g/F_0$.  From left to right, $g/F_0 = 10^{-2}, 3\cdot 10^{-3}, 10^{-3}, 3 \cdot 10^{-4}, 10^{-4}$.  The lines are sigmoid fits as in Fig.~\ref{clogprob}; the fit parameters are plotted in Fig.~\ref{fitparams}.  (b) Clogging probability as a function of $g/F_0$ for fixed values of $w/d$.  From left to right, $w/d = 2.50, 2.25, 2.00, 1.75,$ and 1.40.  The lines are sigmoid fits to the function $P=\{1+\exp[(\ln(g/F_0) - \ln(a))/s] \}^{-1}$ , with centers $g/F_0=a$ having values $1.1\cdot 10^{-3}, 2.8\cdot 10^{-3}, 6.7\cdot 10^{-3}, 1.2\cdot 10^{-2},$ and $1.8\cdot 10^{-2}$ from left to right, and widths $s=0.41, 0.29, 0.28, 0.25,$ and 0.19 respectively.}
    \label{simclog}
\end{figure}

Figure \ref{simclog}(a) shows a similar trend from the simulation data.  Our group's prior work showed that the relevant control parameter is $g/F_0$, the force of gravity normalized by the spring constant acting between the soft particles in the simulation \cite{hong17}; here we keep $F_0=1$.  As gravity is decreased, the particles effectively become harder and it becomes easier to clog; $P_{\rm clog}= 1/2$ moves to larger values of $w/d$.  A different view of simulation data telling the same story is shown in Fig.~\ref{simclog}(b), where now the symbols correspond to fixed values of $w/d$ and $P_{\rm clog}$ decreases as $g/F_0$ is increased.  Narrower hoppers (smaller $w/d$) are easier to clog and thus require higher values of $g/F_0$ to reduce $P_{\rm clog}$.

We wish to quantify and compare the different data sets to understand the influence of particle softness on clogging.  Following Ref.~\cite{hong17}, we fit the $P_{\rm clog}(w/d)$ curves to sigmoidal fits and extract the opening width $w/d$  for which $P_{\rm clog}=1/2$; this characterizes the ability of the system to clog.  To quantify softness, we use the magnitude of deformation $\delta$ a particle has due to its weight, nondimensionalized by the particle diameter $d$.  The experimental deformation $\delta$ is determined by balancing the weight of one particle with the Hertz contact force law, using the $E^*$ modulus data in Table I.  For the simulation data, balancing the gravitational force on a particle with the contact force against a hypothetical horizontal wall leads to $\delta/d = 2g/F_0$.

\begin{figure}[h!]
\includegraphics[width=8cm]{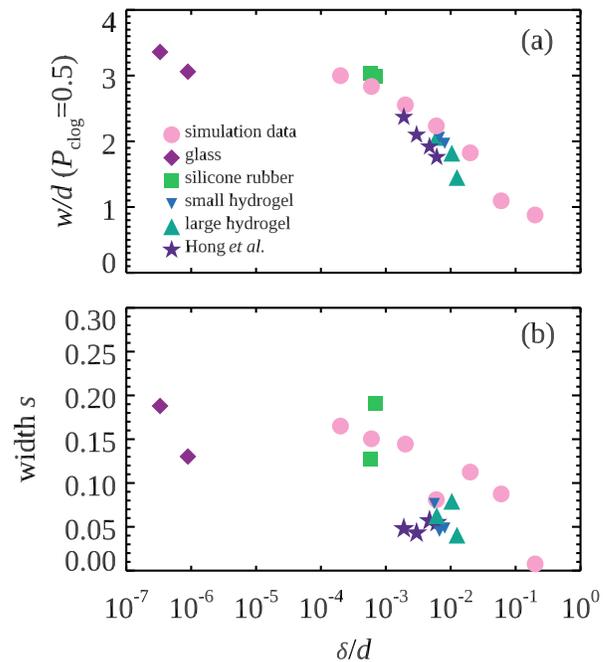}
\centering
\caption{\label{fitparams}
Sigmoidal fit parameters. (a) Centers of sigmoidal fits for different types of particles under varying gravity, with the symbols corresponding to distinct experiments or simulations as indicated in the legend. (b) Width $s$ of sigmoidal fits.  The Hong {\it et al.} data are from Ref.~\cite{hong17} and are from a different set of hydrogel particles.
}
\end{figure}

Figure~\ref{fitparams}(a) shows the parameter $\delta/d$ works fairly well to collapse all of our $w/d(P_{\rm clog}=1/2)$ data, including the laboratory data with hydrogel particles from our group's prior work \cite{hong17}.  The addition of the glass data extends the dynamic range of $\delta/d$ by two orders of magnitude over the prior work.  Considering the differences between the simulation and the experiment, the experimental results are in great agreement with the simulation results suggesting that $\delta/d$ is a good measurement of the importance of softness. As $\delta/d$ gets larger (particles become softer), the hopper opening width needs to become smaller to have a 0.5 clogging probability.  The one unexpected result is that the glass spheres, while being significantly harder and thus at much smaller values of $\delta/d$, still show some slight dependence on $\delta/d$, although nonetheless the glass sphere data are consistent with the overall shape of the curve.  The slight variability of the glass sphere data may be indicating other effects not accounted for in $\delta/d$.  We suspect that the biggest effect is that the larger mass of the glass particles causes the apparatus to vibrate when the particles collide with the walls, and the vibrations may disrupt some arches.  This would be reduced when the tilt angle is reduced (and thus the particles fall slower).  Note that there is one difference between the experiments and simulations:  the experiments use 200 particles, whereas the simulations use 800 particles.  More particles gives more chances to clog \cite{to05,janda08,tang09,lafond13}.  Based on our simulation data, changing the number of particles to 200 would decrease $w/d$ by 0.25 or less in Fig.~\ref{fitparams}(a) (for the circles), which would not qualitatively affect the agreement between experiment and simulation data.

Figure~\ref{fitparams}(b) shows the width of the sigmoidal fit $s$ as a function of $\delta/d$; here the data do not collapse, although they are of somewhat similar magnitude.  Smaller values of $s$ indicate stronger dependence of $P_{\rm clog}$ on $w/d$.  For example, to change $P_{\rm clog}$ from 0.88 to 0.12, $w/d$ needs to increase by $4s$ according to the sigmoidal fit.  All three hydrogel data sets have similar values of $s$, suggesting that the particle type is more influential than $\delta/d$ when it comes to determining $s$.

\subsection{Number of particles remaining in hopper}
\label{secgompertz}

\begin{figure}
\includegraphics[width=8cm]{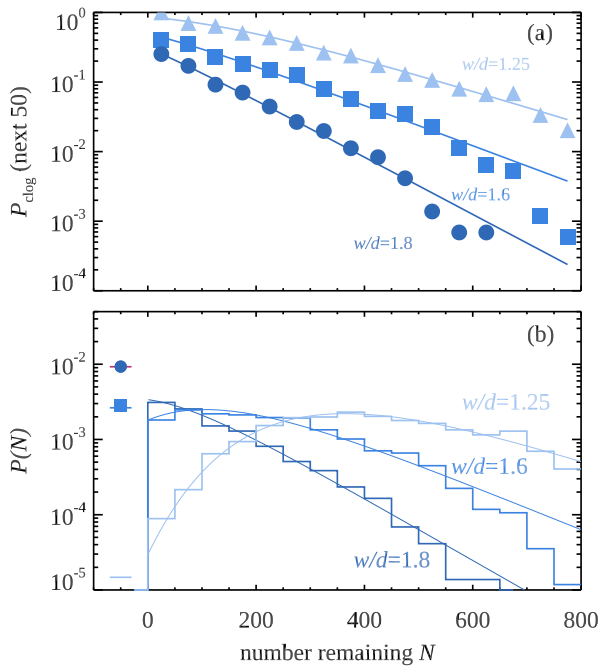}
\caption{\label{hazard}
(a) Hazard rate:  probability of clogging as the next 50 droplets flow out, as a function of the number of particles $N$ left in the hopper.  The lines are the best fit to the Gompertz hazard rate  (integrated over the next 50 droplets).  For  $w/d=1.8, 1.6,$ and 1.25, the data are taken from simulation runs with 1454, 1697, and 1579 trials respectively.  For these data $g/F_0=10^{-2}$; this matches the blue diamonds in Figs.~\ref{simclog}(a) and \ref{gomvsw}(a).
(b) Probability of clogging with $N$ particles left in the hopper.  The lines are the best fit to the Gompertz distribution.  The symbols at the left side of the plot indicate the probability the system does not clog (divided by 50 for comparison to the probability distribution); note that the $w/d=1.25$ data always clogs in the simulation (for 1579 runs) so does not have a symbol.  The horizontal segments for $N<0$ indicate the expected probability that the system does not clog, based on the Gompertz distribution.  The fit parameters are  $w/d=1.25:  h_0=0.040 \pm 0.006, b=0.0055 \pm 0.0003$; $w/d=1.6: h_0=0.014 \pm 0.001, b=0.0067 \pm 0.0003$; $w/d=1.8:  h_0 = 0.0073 \pm 0.0008, b=0.0095 \pm 0.0007$.
}
\end{figure}

We wish to understand how many particles remain in the hopper when it clogs.  A simple hypothesis is that for soft particles, the weight of the particles above a clogging arch matters.  If many particles are still in the hopper, then the arch must bear their weight -- especially in the simulation for which there is no Janssen effect \cite{janssen1895}.  The Janssen effect is a reduction of pressure at the bottom of a container of particles due to friction \cite{nedderman82,shaxby23}, and our simulations do not have friction.  Likewise, recent experiments studying hydrogel particles measured the pressure at the bottom of the hopper and confirmed it depended on how many particles were in the hopper \cite{pongo21}.  Thus, we hypothesize that clogging should be less likely when the hopper is full of particles, and more likely when the hopper has fewer particles.  This is confirmed by the data, shown in Fig.~\ref{hazard}(a).  Here we measure the probability of a clog during the next 50 particles flowing out of the hopper, conditional on not having yet clogged.  For example, all the simulations start with $N=800$ particles.  For Fig.~\ref{hazard}(a) at $N=600$, we are considering all the simulations which did not clog with more than 600 particles, and asking what is the probability that this subset of simulations has a clogging event before reaching $N=550$ particles in the hopper.  This probability rises as the hopper drains (as $N$ decreases), confirming the hypothesis.  The different data sets correspond to different values of $w/d$, with clogging probability larger for the data with smaller $w/d$.  To measure the small probabilities, each data set in Fig.~\ref{hazard}(a) is based on more than 1000 simulations.

The probability measured in Fig.~\ref{hazard}(a) is related to the ``hazard rate,'' the rate of clogging events expected per unit particle exiting the hopper.  Note that unlike probability, the hazard rate is indeed a rate and can be above 1, indicating an extreme likelihood of observing a clog, albeit with a small nonzero chance of not observing a clog.  In contrast, we are focusing on the measured probability, bounded by 1, which behaves conceptually like the hazard rate when $P \ll 1$.  Figure \ref{hazard}(a) is a semilog plot showing that the hazard rate rises exponentially as the hopper drains.  This suggests that the probability distribution of $N$, the number of particles left in the hopper when it clogs, should follow the Gompertz distribution:  the probability distribution corresponding to an exponentially growing hazard rate.  In particular, consider $n=800-N$:  the number of droplets that have flowed out.  If the hazard rate is given by $h(n) = \eta b \exp(b n)$ then $P(n)$ is given by
\begin{equation}
\label{gompertzeqn}
    P(n) = b \eta \exp(\eta+bn) \exp(-\eta e^{bn})
\end{equation}
which is valid for $n, b, \eta \geq 0$.  The cumulative distribution function is given by
\begin{equation}
\label{cume}
   C(n) = 1 - \exp \left( -\eta \left( e^{bn}-1 \right) \right);
\end{equation}
the probability of finding a clog is $P_{\rm clog} = C(n=800)$.
To make more physical sense of the fitting parameters, we define
\begin{equation}
   h_0 = \eta b \exp(800 b)
\end{equation}
so that the hazard rate can be written as 
\begin{equation}
\label{hazardeqn}
    h(N) = h_0 \exp(-b N),
\end{equation}
where $h_0$ has the meaning of the hazard rate as $N\rightarrow 0$ (as the hopper empties).  The parameter $b$ expresses the rapidity of the growth of the hazard rate as the hopper drains.

Figure \ref{hazard}(b) shows the measured probability distribution functions $P(N)$ for the number of particles left in the hopper, and the curved lines show the Gompertz distribution fits to the data (using maximum likelihood) \cite{nelson14}.  The excellent agreement shows that the data are well described by the Gompertz distribution.  The Gompertz distribution fit also predicts the probability that the system does not clog (which is $1-C(n=800)$ from Eqn.~\ref{cume}).  The symbols at the left side of Fig.~\ref{hazard}(b) indicate the observed clogging probability, and the horizontal line segments intersecting the symbols show the predicted clogging probability from the Gompertz distribution fit.  Likewise, the probability of clogging during the next 50 droplets exiting shown in Fig.~\ref{hazard}(a) (symbols) is well fit by the prediction of this quantity from the Gompertz distribution fit (lines).

\begin{figure}
\includegraphics[width=8cm]{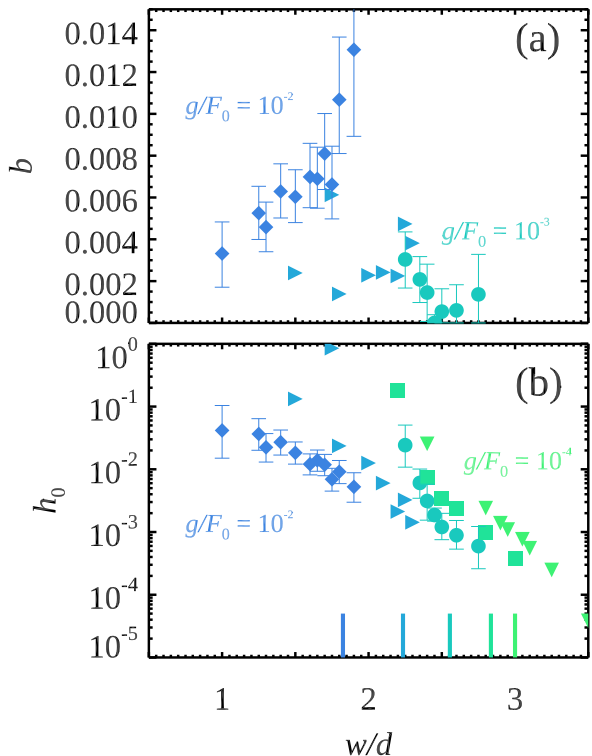}
\caption{\label{gomvsw}
(a) Gompertz distribution fitting parameter $b$ and 
(b) Gompertz distribution fitting parameter $h_0$, as a function of opening width $w/d$.  The data are from the simulation and different symbols and colors correspond to different fixed values of the driving force $g/F_0$.  From left to right, $g/F_0 = 10^{-2}, 3\cdot 10^{-3}, 10^{-3}, 3 \cdot 10^{-4}, 10^{-4}$.  In (a), data are not shown for the two smallest values of $g/F_0$ as $b$ is consistent with zero for those data; see text for discussion.
Representative error bars are drawn for three of the data sets, and represent 90\% confidence intervals.  The symbols are the same in both panels and match the symbols of Fig.~\ref{simclog}(a).  The short vertical lines in (b) indicate the value of $w/d$ for each data set at which $P_{\rm clog}=1/2$, based on the sigmoid fitting shown in Fig.~\ref{simclog}(b).
}
\end{figure}

\begin{figure}
\includegraphics[width=8cm]{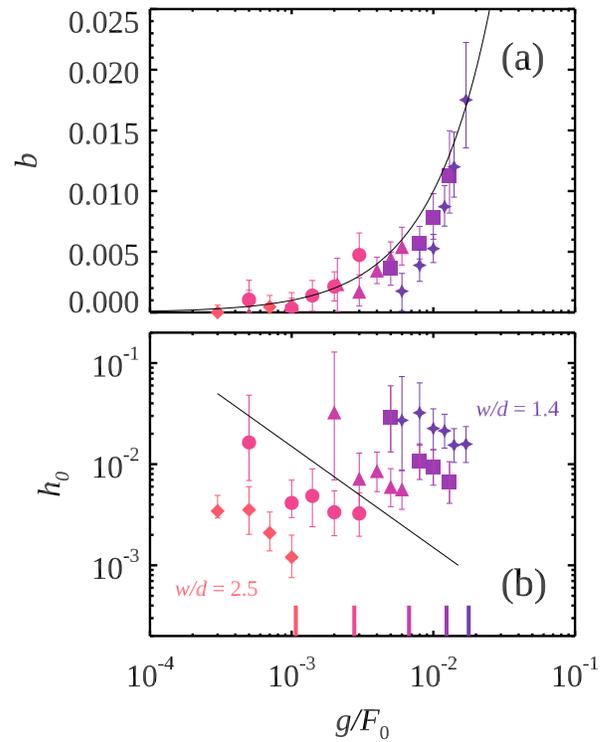}
\caption{\label{gomvsg}
(a) Gompertz distribution fitting parameter $b$.  The black line is the curve $b = g$.
(b) Gompertz distribution fitting parameter $\eta_0$.  The error bars represent 90\% confidence intervals.  The black line has a slope of -1 for comparison.  In both panels, the data are from the simulation and are plotted as a function of $g/F_0$, for fixed values of the opening width $w/d$.  From left to right, $w/d = 2.50, 2.25, 2.00, 1.75,$ and 1.40.  The symbols are the same in both panels and match the symbols of Fig.~\ref{simclog}(b).  The short vertical bars at the bottom of (b) indicate the value of $g/F_0$ for each data set at which $P_{\rm clog}=1/2$, based on the sigmoid fitting shown in Fig.~\ref{simclog}(b).
}
\end{figure}

The influence of $w/d$ and $g$ can be understood by the Gompertz fitting parameters $b$ and $h_0$, shown in Fig.~\ref{gomvsw}.  $b$ stays fairly small, $O(10^{-3} - 10^{-2})$, consistent with needing $O(10^2 - 10^3)$ particles to flow out for significant increases in the hazard rate.  For the simulations with smaller values of $g/F_0$, the hazard rate does not measurably increase as the hopper drains; it would require significantly more data to detect the slight increase.  Accordingly, $b\approx 0$ within our uncertainty and those data are not plotted in Fig.~\ref{gomvsw}(a) for $g/F_0 < 10^{-3}$.  In Fig.~\ref{gomvsw}(b), $h_0$ decreases dramatically as $w/d$ increases, indicating that the system becomes increasingly unlikely to clog even as $N\rightarrow 0$.
Note that the error bars in Fig.~\ref{gomvsw} are largest when $P_{\rm clog} \rightarrow 1$ or $\rightarrow 0$, for which there is less variability in the observations of $N$ and thus fewer constraints on the fitting.

% In contrast, as $w/d$ is decreased, $\eta_0$ rises to well above 1, reflecting the observation that not only do we always see clogging in the simulation, but that it always is observed to clog with many particles still remaining in the hopper -- which is illustrated by the $w/d=1.25$ data in Fig.~\ref{hazard}(b) (lightest color curve) for which the system did not empty out for any of the 1579 simulation runs. 

A complementary view of the simulation data is given in Fig.~\ref{gomvsg}, where each symbol type corresponds to a fixed value of $w/d$, and the horizontal axis shows the dependence on $g$; the corresponding clogging probability data are shown in Fig.~\ref{simclog}(b).  If all that mattered for the hazard rate (at fixed $w/d$) is the weight of the pile above for a given $N$, then it would make sense that $b \sim g$, see Eqn.~\ref{hazardeqn} noting the weight is $\sim N g$.  The prediction $b = g$ is the black curve drawn in Fig.~\ref{gomvsg}(b), showing rough qualitative agreement for all of the data, although underestimating the results for large $w/d$ (red symbols on left side of graph) and overestimating for small $w/d$ (purple symbols on right side of graph).  This predicts that $b$ is constant when considering data at constant $g$, whereas Fig.~\ref{gomvsw}(a) that $b$ varies by about a factor of 3 at fixed $g/F_0$ and varying $w/d$, further evidence that $b \sim g$ is only roughly true.
The data of Fig.~\ref{gomvsg}(b) shows that $h_0$ decreases roughly as a power law with $g/F_0$.  Fitting $h_0 \sim g^{-\alpha}$, the data give $\alpha = 1.0 \pm 0.5$; a line with slope -1 is shown for comparison in Fig.~\ref{gomvsg}(b).

The limit $g\rightarrow 0$ is important in that it represents perfectly hard particles \cite{arevalo16}.  Our group's prior work suggested that the data of Fig.~\ref{fitparams}(a) should reach an asymptote for small $g$, although as noted above the glass particles data suggest that there is still some additional dependence on the forcing, perhaps due to vibrational effects.  The Gompertz distribution fit parameter $b$ is the ``rate of change of the hazard rate'' [recall the hazard rate is given by $h(n) = \eta b \exp(b n)$].  Thinking just of the simulation data, Fig.~\ref{gomvsg}(b) suggests that if $b \sim g$, the hard particle limit is $b \rightarrow 0$ (holding the product $\eta b$ constant as the limit is taken), signifying that the hazard rate is independent of the number of particles in the hopper.  This certainly seems to be the case for the classic clogging of hard particles, for which the output flux is independent of the number of particles in the hopper \cite{beverloo61,aguirre10,thomas15}.  For the fit parameter $h_0$, assuming $b=0$ for hard particles means that $h_0$ is the constant hazard rate for clogging, which should not depend on $g$ for hard particles.  This is for example consistent with an experiment that used a centrifuge to vary $g$, finding no dependence on $g$ \cite{dorbolo13}, although Ref.~\cite{hong17} pointed out that those particles were not infinitely hard but rather had $\delta/d \approx 10^{-6}$.

To compare to the experiment, when the experiment clogs we measure the number of particles remaining in the hopper.  For the experiment, this is not an exact count but rather an approximation based on image analysis.  Ideally we would wish to use image analysis to directly identify and count the particles.  However, given that many of the experiments use soft particles which are deformed and closely touching, this sort of image analysis proved problematic.  Instead, when the hopper clogs, we determine the boundary of the region of the image containing particles in the hopper, and then measure the area contained within that region.  To calibrate this we took 10 photographs with exactly 200 particles in the hopper for the same imaging conditions as the experiment (same particle type, same tilt angle, same lighting conditions) and measure the area those 200 particles occupy.  This then gives us the mean area per particle.  For the experimental data of interest, our uncertainty in number of particles remaining in the hopper is $\pm 10$ from this method:  as will be seen, the results do not depend sensitively on this uncertainty.  We used this method to estimate the number $N$ of particles left in the hopper when samples clogged, using 300 trials for three different particle types, and $w/d$ such that $P_{\rm clog} \approx 0.7$ so that a large number of clogging events would be observed.

\begin{figure}[h!]
\includegraphics[width=8cm]{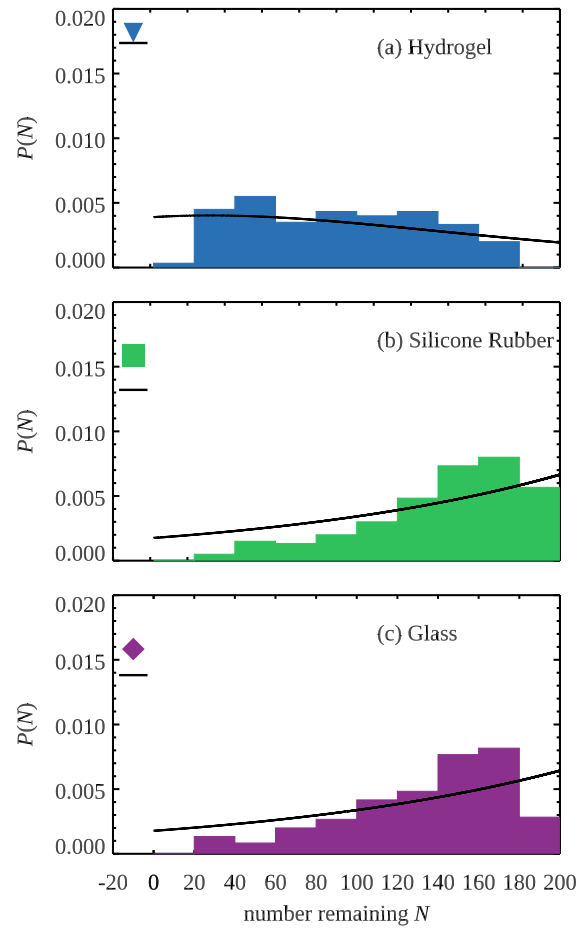}
\centering
\caption{\label{gom2}
Histograms of probability of clogging with $N$ particles left in the hopper for (a) small hydrogel particles ($w/d=2.03$), (b) silicone rubber particles ($w/d=3.03)$, and (c) glass particles ($w/d=2.91$).  All data are taken at tilt angle $\theta = 50^\circ$.  The solid black lines are the best fit to the Gompertz distribution; note that this is a reasonable fit in (a) and a less reasonable fit in (b,c), showing that the latter data are not consistent with a Gompertz distribution. The symbols at the left side of the plot indicate the probability the system does not clog (divided by 20 for comparison to the probability distribution). The horizontal segments for $N < 0$ indicate the expected probability that the system does not clog, based on the Gompertz distribution.  The fitting parameters are (a) $h_0=0.011 \pm 0.004, b=0.009 \pm 0.003$; (b) $h_0=0.007\pm 0.1, b\approx 0$; (c) $h_0=0.006\pm 0.1, b\approx 0$.
}
\end{figure}
% See notes.hopper4d 6-23-2021.  Confirmed on 7-14-2021 that I did calculate eta_0.  Revised on 10-8-2021 with gompertz6expt (new fitting)

The experimental data are plotted in Fig.~\ref{gom2}, along with the best fit to the Gompertz distribution.  The hydrogel data are reasonably well fit, the silicone rubber and glass data are less well fit.  For the latter two cases, the fit significantly underpredicts the probability of not clogging.  Additionally, for a sample that does not clog $30\%$ of the time, one expects that when clogging occurs, it should more often occur with fewer particles in the hopper.  That is the argument given above for the simulation data, that when the hopper drains out the pressure decreases and thus it is easier to form an arch that can support the weight of the remaining particles.  However, for the silicone rubber particles and the glass particles, it appears that it is most likely for the experiment to clog near the start of the experiment; and the more particles that have flowed out, the less likely it is to clog.  We can speculate as to the causes.  First, the silicone rubber particles have more friction, whereas the glass particles are significantly harder; both of these may frustrate the argument about pressure making a difference to the clogging arch formation.  Second, the particles also are more massive, and it may be that as they fall though the hopper, they add extra vibrations to the apparatus.  Vibrations are well known to destabilize clogging arches \cite{zuriguel14}.  Perhaps it is easier for an arch to form before the apparatus starts shaking too much, and harder after particles are flowing out in significant quantity.  In any case, we note that by comparison the hydrogel data agree reasonably well with the Gompertz distribution fit.  It is further interesting to note that despite the imperfect Gompertz distribution fit, nonetheless the silicone rubber and glass particles fit well on our clogging plot in Fig.~\ref{fitparams}(a).

To measure the quality of the Gompertz distribution fits (for both simulation and experimental data) we use the Kolmogorov-Smirnov test \cite{numrecipes}.  This test compares the observed cumulative distribution function $C_{\rm obs}(n)$ and compares it to the predicted $C(n)$ from Eqn.~\ref{cume}.  The K-S statistic is $D = \max( |C_{\rm obs}(n) - C(n)| )$; smaller is better, although this depends on how much data one has.  For the three simulations with more than 1000 trials (Fig.~\ref{hazard} data), $D < 0.028$ and the probability of this being a good fit is relatively high ($P = 0.28 - 0.86$) \cite{numrecipes}.  For the other simulations with 100 trials, we find $D \sim 0.01 - 0.10$ and generally the probability of a good fit remains high ($P=0.13 - 0.99$).  For the hydrogel particles $D=0.07$, $P=0.28$; for silicone rubber, $D=0.11, P=0.012$; and for glass, $D=0.08, P=0.11$.  While the hydrogel fit isn't as good as one might wish, it is better than the rubber and glass data, in agreement with the qualitative behavior of Fig.~\ref{gom2}.  As a reminder, the fitting is done with the maximum likelihood method which does not inherently have a quality of fit measure, and thus is not optimizing the probability based on the K-S statistic.

\subsection{Arch size}

\begin{figure}[h!]
\includegraphics[width=8cm]{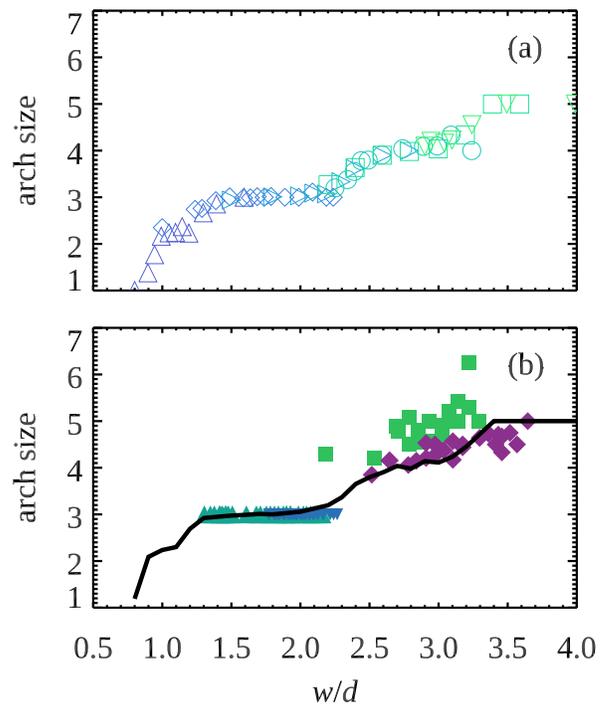}
\centering
\caption{\label{archsize}
Average arch size as a function of $w/d$, the ratio of the hopper exit width $w$ to the droplet diameter $d$. (a) Data from the simulation. Different values of gravitational driving are indicated by the different symbols and colors; from left to right, $g/F_0$ = 0.03, 0.01, 0.003, 0.001, 0.0003, and 0.0001.  The symbol shape and color matches those shown in Fig.~\ref{simclog}(a).  (b) The line reprises the simulation data from (a).  The symbols are the experimental data using different types of particles under the influence of different values of gravity.  The symbols are the same as Fig.~\ref{fitparams}; in particular, the green squares that are outliers correspond to the silicone rubber particles, which have a significantly higher coefficient of static friction.
}
\end{figure}

When a clogging event occurs, we count the number of particles forming the arch. We then average this over all clogging events for a given condition.  The mean arch size simulation results are shown in Fig~\ref{archsize}(a); different symbols and colors indicate different values of the gravitational driving $g$.  The excellent collapse of the data shows that the average arch size solely depends on the exit width $w/d$ independent of the magnitude of $g$.  The smallest arch has one particle, and this is only seen for small values of $w/d < 1.0$.  As $w/d$ increases to 1.5, the average arch size increases to three. There is a plateau for $1.5 \lesssim w/d \lesssim 2.3$ where the average arch size is constant at 3.  This plateau was also seen in prior experimental data with hard particles \cite{lopezrodriguez19}, although in that work it was more pronounced when the hopper wedge angle was steeper than our moderate $34^\circ$ angle.  At higher values of $w/d$ there is a smaller plateau with arch size equal to 4, and then a bit of data with the mean arch size rising to 5 at the lowest value of $g$ ($g=10^{-4}$) and $w/d \gtrsim 3.5$.

To compare the experiment and simulation, the simulation data are replotted in Fig.~\ref{archsize}(b) as the line, and the experimental results are plotted with symbols [corresponding to the legend in Fig.~\ref{fitparams}(a)].  The experimental results for the hydrogel particles (blue triangles) and glass particles (purple diamonds) agree with the simulation result supporting that the arch size is affected by the opening width and is independent of the magnitude of gravity. This strongly suggests for these particles -- ranging from quite soft to quite hard, more than 5 orders of magnitude in $\delta/d$ -- the clogging arch is solely determined by geometry.  In contrast, the results for silicone rubber particles (green squares) deviate significantly from the other results:  for a given opening width, the average arch size will be larger than that for simulation and the glass particles. We attribute the difference between the silicone rubber particles and the other data to be due to the silicone rubber particles’ large coefficient of sliding friction. The sliding friction for silicone rubber particles is $0.4\pm 0.2$, while the sliding friction for glass particles is $0.009\pm 0.002$, and the simulation has no friction.  Indeed, the arch shown in Fig.~\ref{ranpics}(d) has one particle that is clearly held in place by friction.  While this is an uncommon result, this clear frictional effect is observed several times in our experiments with the silicone rubber particles (and never with any other particles).

\section{Conclusions}

Work in recent years has shown that the clogging of soft particles is qualitatively different than hard particles \cite{hong17,ashour17,stannarius19,harth20,pongo21}.  We extend this prior work with two new types of particles, silicone rubber and glass, showing that there is a relatively continuous transition from the softest particles to the hardest [Fig.~\ref{fitparams}(a)].  The agreement between hydrogel data, silicone rubber data, glass data, and simulation data -- with no free fitting parameters -- is strong evidence for universal behavior of soft particle clogging.  This is further supported by examining the mean arch size, which is a function of only $w/d$ (exit opening width $w$ divided by the particle diameter $d$).  The caveats are that this is not true for the silicone rubber particles for which friction is nearly two orders of magnitude higher; and the mean arch size is also known to depend on the hopper wedge angles \cite{lopezrodriguez19}.  We also analyze the number of particles left in the hopper, further finding a difference between the silicone rubber and glass particles, as compared to the simulation and hydrogel particles.  For silicone rubber and glass particles, if they clog, it is slightly more likely to do so near the start of the experiment.  In contrast, for the hydrogel and simulated particles, they are more likely to clog near the end of the experiment, which serves as strong evidence that the hydrostatic pressure of the particles in the hopper \cite{pongo21} breaks arches and prevents clogging.

Our work shows that for soft hydrogel particles and simulated soft particles, the number of particles left in the hopper when a clog occurs is well fit by the Gompertz distribution.  This distribution applies when the clogging ``hazard rate'' rises exponentially as the hopper drains.  The direct implication is that the hydrostatic pressure of the soft particles influences clogging -- that it is harder to form a clogging arch with many particles in the hopper as the hydrostatic pressure causes the soft particles in the arch to deform and break the arch.  A subtler implication of the Gompertz distribution fit is that there is some chance of the hopper clogging even when the hopper is full of particles; albeit that the chance is exponentially small.  This further suggests that even with large hopper openings $w/d$ there is still some chance of clogging,  consistent with prior observations of clogging that suggested that there is no critical exit size for causing clogging;  rather, clogging becomes exponentially unlikely as the opening size is increased \cite{to05,janda08,thomas15}.  Finally, the Gompertz distribution fit also implies that in the opposite limit of a small opening size, there may nonetheless be some finite probability that the hopper does {\it not} clog.  It seems plausible that there is some limit on this, that for opening widths smaller than the particle size and sufficiently stiff particles, the system will always clog.  Nonetheless, the results imply that the ability to completely flow out may persist to surprisingly narrow exit openings, even if the chance to not clog becomes exponentially rare.

\section{Acknowledgements}

This material is based upon work supported by the National Science Foundation under Grant No. (CBET-2002815).  We thank I.~Nemenman for bringing the Gompertz distribution to our attention.  We thank Y. Cheng, S. Franklin, C. Lerch, M. Morrell, C. O'Hern, M. Shattuck, and J. Treado for  helpful discussions.

\end{document}